\documentstyle[12pt]{article}
\def\ba{\begin{eqnarray}}
\def\ea{\end{eqnarray}}

\def\lb{\label}
\def\be{\begin{equation}}
\def\ee{\end{equation}}

\begin{document} \title{Massless geodesics in $AdS_5\times Y(p,q)$ as a superintegrable system} \author{Emilio
Rub\'in de Celis\thanks{Departamento de Física, Facultad de Ciencias Exactas y Naturales, Universidad de Buenos Aires and IFIBA, CONICET, Cuidad Universitaria, Buenos Aires 1428, Argentina. e-mail: erdec@df.uba.ar}$\,$ and Osvaldo P. Santill\'an \thanks{CONICET-Instituto de
Investigaciones Matem\'aticas Luis Santal\'o, Universidad de Buenos Aires. e-mail: firenzecita@hotmail.com and
osantil@dm.uba.ar.} } \date {} \maketitle \begin{abstract} A Carter like constant for the geodesic motion in the
$Y(p,q)$ Einstein-Sasaki geometries is presented. This constant is functionally independent with respect to the
five known constants for the geometry. Since the geometry is five dimensional and the number of independent
constants of motion is at least six, the geodesic equations are superintegrable. We point out that this result
applies to the configuration of massless geodesic in $AdS_5\times Y(p,q)$ studied by Benvenuti and Kruczenski
\cite{benve}, which are matched to long BPS operators in the dual N=1 supersymmetric gauge theory.
\end{abstract}

\section{Introduction}

The present work deals with a superintegrable problem. Roughly speaking, a mechanical system is called
superintegrable if the number of its functionally independent constants of motion is larger than its number of
degrees of freedom.

The classical example of a superintegrable system is the Kepler one. As is well known, for the motion in a
generic central field, the energy $E$ and the component of the angular momenta perpendicular to the plane of
motion $L_z$ are conserved. Since the motion takes place in a plane, all these problems are integrable. For the
Kepler problem, there is a further conserved quantity namely, a component of the Runge-Lenz vector. The set of
these three constant of motion is functionally independent, therefore the Kepler problem is superintegrable.
Similar considerations apply to the central harmonic oscillator in three dimensions. For both systems, the
closed trajectories are ellipses.

The maximal number of functionally independent constant of motion that a mechanical system with $n$ degrees of
freedom may admit is $2n-1$. Systems possessing this number of constants of motion are known as maximally
superintegrable. For $n=2$ a maximal superintegrable system admits three constants of motions. Thus the Kepler
problem is maximally superintegrable.

Superintegrable systems are gifted with special properties, some of them are intrinsic and some others depends
on the problem under consideration. For the Kepler motion, some interesting features emerge when upon
quantization. The Runge-Lenz vector becomes, by the correspondence principle, an operator which commutes with
the hamiltonian of the particle. The algebra constituted by the hamiltonian, the angular momentum and the
Runge-Lenz vector is not closed, in fact is an infinite dimensional twisted loop algebra \cite{daboul}. But when
restricted to subspaces of constant negative energy, which correspond to bound states, the resulting symmetry
algebra is isomorphic to $SO(4)$. Thus the expected symmetry group for a central field, which is $SO(3)$, is
enhanced for the Newton potential to $SO(4)$. This enhancement explains the accidental degeneration of the
energy levels of the hydrogen atom, i.e, the independence of the energy levels with respect to the total angular
momenta of the particle. In fact Pauli, Bargmann and Fock \cite{pauli}-\cite{pauli3} have shown that, due to the
presence of the Runge-Lenz operator, it is possible to obtain the bound state spectra of the hydrogen without
solving the Schrodinger equation explicitly. This suggest that the discovery of superintegrability for a given
problem may dramatically simplify the study of its properties.

 It was Sommerfeld who pointed out that if, for a given potential, the Hamilton-Jacobi equation is separable in
 more than one coordinate system, then the problem is superintegrable \cite{sommerfeld}. This statement was
 extensive studied by Smorodinskyi, Winternitz and collaborators, who were able to classify the potentials in
 two dimensions which are separable in more than one coordinate system \cite{winternitz1}-\cite{winternitz2}.
 For three dimensional flat space, it was Eisenhart who classified the possible coordinates systems for which
 separation takes place, together with the form of the potentials that permits such separation
 \cite{Eisenhart1}-\cite{Eisenhart2}. Further investigation was made in \cite{winternitz3} where the three
 dimensional potentials that are separable in spherical coordinates and one more coordinate system were found.
 The remaining possibilities in three dimensions were classified in \cite{evans} later on.

The achievements described above motivated intense research in the subject. In recent years several integrable
systems were found, some examples are in \cite{ballesteros1}-\cite{gibbons17} and references therein. These
examples consist in a wide variety of physical systems such as the Kepler problem in arbitrary dimensions and
its extensions in presence of magnetic monopoles, and generalizations of known systems to spaces of non zero
curvature in several dimensions.

In the present work it will be shown that the equations for the geodesic motion over the Einstein-Sasaki metric
defined on the $Y(p,q)$ manifolds discovered in \cite{Martelli sparks}-\cite{Gauntlett:2004yd} are
superintegrable. The main technical tool for obtaining this result are  Killing and Killing-Yano tensors
\cite{Yano}. These tensors play a significant role for the integrability of the geodesic equations in the
rotating black hole background \cite{CarterKerr}-\cite{Kinnersley}. Several work related to this topic were
reviewed in \cite{yo}, and more recent references are \cite{nuevo1}-\cite{nuevo4}.

The present text is organized as follows. In section 2.1 the main features of Einstein-Sasaki manifolds and
Calabi-Yau cones are briefly reviewed, together with a description of the $Y(p,q)$ geometries. In section 2.2
the defining equations for the configurations of massless geodesics on $AdS_5\times Y(p,q)$ considered in
\cite{benve} are shown to be integrable. The material in section 2 is, of course, not new. In section 3.1 the
main features of Killing and Killing-Yano tensors as generators of hidden symmetries are reviewed. These tools
are applied in section 3.2 to show that the configurations of massless geodesics mentioned above admit a further
constant of motion which is functionally independent with respect to the ones found in \cite{benve}. This is
checked explicitly, and it is concluded that the configuration of massless geodesics in the geometry is
superintegrable. In section 4 some consequences related this hidden symmetry are derived. It is found that the
superparticle worldline action \cite{Superparticula1}-\cite{Superparticula4} with the $Y(p,q)$ geometries as
space-time metric admit exotic supersymmetries. Additionally, it is found that this symmetry is not anomalous,
in the sense that it corresponds to an operator which commutes with the laplacian defined over the $Y(p,q)$
geometry. Properties of the Dirac operator are also briefly commented. In section 5, some open perspectives and
future lines of work are discussed. Although the validity of our results are checked in the text, some
mathematical statements which were used for obtaining them are collected in the appendix. This is in order to
separate the statement of the results from the description of how they were obtained, for sake of clarity.

\section{Preliminar material} \subsection{A brief description of the Einstein-Sasaki metrics on $Y(p,q)$}

Since this work is related to Einstein-Sasaki metrics, it will be convenient to give a brief description of
their main properties. Einstein-Sasaki metrics are directly connected to non compact Calabi-Yau cones. Recall
that a non compact Calabi-Yau metric $g$ is by definition a $2n$ dimensional one defined over a space $M_{2n}$
and whose holonomy is $SU(n)$ or a subgroup of $SU(n)$. All these metrics are in particular Ricci flat, and
there exist always a local choice of the basis $e^a$ for which the metric takes the diagonal form $$
g=\delta_{ab}e^a\otimes e^b, $$ and for which the symplectic two form \be\lb{simp} \omega=e^1\wedge
e^2+..+e^{2n-1}\wedge e^{2n}, \ee and the $(n,0)$ form \be\lb{complex} \Omega=(e^1+ie^2)\wedge...\wedge
(e^{n-1}+ie^{n}) \ee are closed. The converse of these statements are also true, that is, a non compact metric
satisfying the conditions enumerated above is Calabi-Yau. Manifolds for which $d\omega=0$ are sympletic. The
closure of $\Omega$ implies that the almost complex structure $J$ defined by $\omega(\cdot, \cdot)=g(\cdot,
J\cdot)$ is integrable and thus, the manifold is complex. Complex sympletic manifolds of this type are Kahler
and therefore, any Calabi-Yau metric is automatically Kahler. In fact, a Ricci flat Kahler metric is locally
Calabi-Yau. Details of these assertions can be found in standard books on the subject \cite{books}.

The relation between Einstein-Sasaki manifolds and Calabi-Yau cones is as follows. Consider the family of $2n$
dimensional cones given by \be\lb{cone} g_{2n}=dr^2+r^2 g_{2n-1}, \ee with metric $g_{2n-1}$ which does not
depends on the coordinate $r$ and is defined over a $2n-1$ manifold. The distance element (\ref{cone}) is
singular at $r=0$ unless $g_{2n-1}$ is the canonical metric on the sphere $S^{2n-1}$. If the cone (\ref{cone})
is Calabi-Yau, then $g_{2n-1}$ is known as Einstein-Sasaki. This relation can be taken as the definition of an
Einstein-Sasaki metric, in a local sense. In fact, since a Calabi-Yau cone is Ricci flat, the metric $g_{2n-1}$
is Einstein with non zero cosmological constant. The converse of these statements are all true, that is, any
Einstein-Sasaki metric defines a Calabi-Yau cone by (\ref{cone}). Several properties for these metrics are
collected in the appendix but were extensively reviewed in \cite{Boyer}-\cite{Boyer2}.

The Einstein-Sasaki metrics which we will be concerned with are the ones defined over the $Y(p,q)$ manifolds
\cite{Martelli sparks}-\cite{Gauntlett:2004yd}. There exist a local coordinate system for which the distance
element takes the following form $$ g_{p,q}= \frac{1-y}{6} (d\theta^2+\sin^2\theta d\phi^2)+ \frac{dy^2}{6 p(y)}
+ \frac{q(y)}{9} (d\psi - \cos\theta d\phi)^2 $$ \be\lb{metric2} + w(y)[d\alpha+f(y)(d\psi-\cos\theta d\phi)]^2
 \ee
with $$ w(y)= 2\frac{a - y^2}{1 - y},$$ \be\lb{wqfdef} q(y)=\frac{a-3 y^2 + 2 y^3}{a-y^2},\ee $$
f(y)=\frac{a-2y+y^2}{6(a-y^2)} ,$$ and \be\lb{pdef} p(y) = \frac{w(y) q(y)}{6} = \frac{a-3y^2+2y^3}{3(1-y)} \ee
The coordinates $(\theta,\phi,y,\alpha,\psi)$ take values in the range
  \be \lb{arange}
  0\leq \theta \leq \pi, \qquad 0\leq \phi \leq 2\pi , \qquad y_1\leq y \leq y_2, \ee
  $$
 0\leq \alpha \le 2\pi l , \qquad 0\leq \psi \leq 2\pi.
  $$
 The constant $a$ appearing in the metric and the constants $y_{1,2}$ and $l$ determining the interval of the
 values of the coordinates can be expressed in terms of two integers $p$ y $q$ defining the manifold
$$ a=3y_1^2-2y_1^3, $$ $$ l=\frac{q}{3q^2-2p^2+p\sqrt{4p^2-3q^2}}, $$ $$ y_{1,2} = \frac{1}{4p}\left(2p \mp 3q
-\sqrt{4p^2-3q^2}\right). $$ The constants $y_{1,2}$  are the zeroes of the function $p(y)$ appearing in the
expresion for the metric. In addition, there is a third zero for $p(y)$ given by $y_3=\frac{3}{2}-y_1-y_2$. A
global analysis of these metrics can be found in \cite{Gauntlett:2004yd}.

The coordinate change $\alpha=-\frac{\beta}{6}-\frac{\psi'}{6}$, $\psi=\psi'$ take the distance element
(\ref{metric2}) to the following form \be\lb{ypq} g_{p,q} = \frac{1-y}{6} (d\theta^2+\sin^2\theta d\phi^2)+
\frac{dy^2}{6 p(y)} + \frac{1}{36} q(y)w(y)(d\beta + \cos\theta d\phi)^2 \ee $$ + \frac{1}{9}[d\psi'-\cos\theta
d\phi+y(d\beta+\cos\theta d\phi)]^2, $$ Note that this expression is of the form \be\lb{loco}
g_{p,q}=\frac{1}{9}(d\psi'+A)^2+g_4, \ee with the 1-form $A$ given by \be\lb{1f} A=-\cos\theta
d\phi+y(d\beta+\cos\theta d\phi), \ee and the four dimensional metric $g_4$ given by \be\lb{ke}
g_4=\frac{1-y}{6} (d\theta^2+\sin^2\theta d\phi^2)+ \frac{dy^2}{6p(y)} + \frac{1}{36} q(y)w(y)(d\beta +
\cos\theta d\phi)^2. \ee From (\ref{loco})-(\ref{ke}) it follows that the vector field $V=\partial_{\psi'}$ is
Killing. The form (\ref{loco}) is quite general in the theory of Einstein-Sasaki manifolds
\cite{Boyer}-\cite{Boyer2}. The Killing vector field is known as the Reeb vector field and the four dimensional
metric is in general Kahler Einstein, with Kahler form $\omega=dA$. Einstein-Sasaki metrics are locally $U(1)$
fibrations over a Kahler-Einstein manifolds in general (see appendix).

\subsection{Massless strings in $AdS_5\times Y(p,q)$ as an integrable system}

The $Y(p,q)$ geometries described in the previous subsection are relevant in the context of the AdS/CFT
correspondence \cite{Malda}. For instance, the study of semiclassical strings in backgrounds of the form
$AdS_{5}\times Y(p,q)$ with the local distance element \be\lb{metric1} g_{10}= - dt^2\, \cosh^2\!\rho + d\rho^2
+ \sinh^2\!\rho\, d\Omega_3^2 + g_{p,q}, \ee together with their conserved quantities gives information about
the anomalous dimensions of certain $N=1$ supersymmetric gauge theory, by the gauge/gravity duality
\cite{Kruczenski}. A particular configuration of interest is given by the non massive geodesics in the reduced
metric \be\label{redgeod} g = -dt^2 + ds_{p,q}^2 = -dt^2 + g_{ab} dx^a dx^b, \ee which describes a particle like
limit of the strings. In (\ref{redgeod}), $g_{ab}$ denotes the metric $Y(p,q)$ described in (\ref{metric2}), $t$
is the global time coordinate in $AdS_{5}$, the non massive point like string is located in $\rho=0$ and the
movement takes place in the internal space $Y(p,q)$. The action for such particle limit configuration is
 \be\lb{kru}
 S = \frac{\sqrt{\lambda}}{2} \int d\tau \left( -\dot{t}^2 + g_{ab} \dot{x}^a \dot{x}^b \right)
 \ee
where $\sqrt{\lambda}=(R/l_s)^2$ is the effective string tension. The equations of motion should be supplemented
with the null geodesic constraint \be
 -\dot{t}^2 + g_{ab} \dot{x}^a \dot{x}^b = 0. \label{constraint}
 \ee
The Euler-Lagrange equation for $t$ gives that $t=P_t\,\tau$ with $P_t$ the conjugate momenta of $t$. $P_t$ is
then constant and represent the energy of the string configuration. In these terms the action is reduced to $$
 S =\int d\tau L = \frac{\sqrt{\lambda}}{2} \int d\tau (g_{ab} \dot{x}^a \dot{x}^b)
 $$
 $$
=\frac{\sqrt{\lambda}}{2} \int d\tau \bigg\{ \frac{1-y}{6} (\dot{\theta}^2 + \sin^2\theta\dot{\phi}^2) +
\frac{1}{w(y)q(y)} \dot{y}^2+ \frac{q(y)}{9} (\dot{\psi}-\cos\theta\dot{\phi})^2 $$ \be\lb{acES} +w(y)
[\dot{\alpha} +f(y)(\dot{\psi}-\cos\theta\dot{\phi})]^2\bigg\}, \ee which describes a free particle in the
Einstein-Sasaki geometry. The conjugate moments are by definition \be P_a = \frac{\partial L}{\partial
\dot{x}^a} \ee and in terms of these moments the Hamiltonian is expressed as \be H = \frac{1}{2} g^{ab} P_a P_b.
\ee It can be seen from the isometries of (\ref{metric2}) that the quantities $P_{\phi}$, $P_{\psi}$ y
$P_{\alpha}$ are conserved. Additionally, the square of the $SU(2)$ angular momenta \be\lb{J2} J^2 =
P_{\theta}^2 + \frac{1}{\sin^2\theta} \left(P_{\phi}+ \cos\theta P_{\psi}\right)^2 + P_{\psi}^2, \ee is also
conserved. The full set of momenta can be expressed in terms of the velocities as follows \be\lb{1}
\frac{1}{\sqrt{\lambda}}P_y = \frac{1}{6 p(y)} \dot{y}, \ee \be\lb{2} \frac{1}{\sqrt{\lambda}}P_{\theta} =
\frac{1-y}{6} \dot{\theta}, \ee \be\lb{3} \frac{1}{\sqrt{\lambda}}\left(P_{\phi} + \cos\theta P_{\psi}\right)=
\frac{1-y}{6} \sin^2\theta \dot{\phi},  \ee \be\lb{4} \frac{1}{\sqrt{\lambda}}\left(P_{\psi} - f(y) P_{\alpha}
\right)= \frac{q(y)}{9} \left(\dot{\psi} - \cos\theta \dot{\phi}\right), \ee \be\lb{5}
\frac{1}{\sqrt{\lambda}}P_{\alpha}=w(y) \left(\dot{\alpha} + f(y)  \left(\dot{\psi} - \cos\theta
\dot{\phi}\right)\right) \ee and in these terms the Hamiltonian may be expressed as \be\lb{6} 2\lambda H=
\lambda\kappa ^2= \frac{1}{2} 6 p(y) P_{y}^2 + \frac{6}{1-y} \left(J^2-P_{\psi}^2\right)        +
\frac{1-y}{2(a-y^2)} P_{\alpha} ^2 \ee $$ +\frac{9(a-y^2)}{a-3y^2+2y^3}\left(P_{\psi}-
\frac{a-2y+y^2}{6(a-y^2)}P_{\alpha} \right)^2. $$ In the last equation formula (\ref{constraint}) has been taken
into account, in order to  related $\kappa$ with $H$. Thus there are five functionally independent conserved
quantities for the problem namely $P_\phi$, $P_\psi$, $P_\alpha$, $J^2$ y $H$ and the equations defining the
problem constitute an integrable system. The purpose of the following sections is to present a further conserved
quantity which is functionally independent with respect to these. The presence of this quantity means that the
problem is superintegrable.

\section{Superintegrability of the massless strings in $AdS_5\times Y(p,q)$}

The main technical tool for the following discussion will be Killing and Killing-Yano tensors. We review their
role for finding constants of motion for particle actions such as (\ref{kru}). After this brief review, we
present explicit Killing and Killing-Yano tensors for the $Y(p,q)$ geometries. This result will imply that the
geodesic equations for this geometry is a superintegrable system. We leave for the appendix the technical
details for the construction, for sake of clarity.

\subsection{Killing and Killing-Yano tensors}

The motion of a free particle on a geometry $(M, g_{\mu\nu})$ takes place along a geodesic. The set of geodesics
for the geometry is described by the following action \be\lb{lagro}
S=\int_{\tau_0}^{\tau_1}\textit{L}\,d\tau=\int_{\tau_0}^{\tau_1}\frac{1}{2}\,g_{\mu\nu}(x)\dot{x}^\mu\dot{x}^\nu
d\tau, \ee in particular (\ref{kru}) is of this form. The variation of (\ref{lagro}) with respect to
infinitesimal transformations of the trayectory $\delta x$ and $\delta \dot{x}$ is \be\lb{lagro2} \delta
S=\int_{\tau_0}^{\tau_1}\bigg[\frac{\delta \textit{L}}{\delta x^\mu}-\frac{d}{d\tau}\bigg(\frac{\delta
\textit{L}}{\delta \dot{x}^\mu}\bigg)\bigg]\delta x^\mu
d\tau+\int_{\tau_0}^{\tau_1}\frac{d}{d\tau}\bigg(\frac{\delta \textit{L}}{\delta \dot{x}^\mu} \delta x^\mu\bigg)
d\tau \ee $$ =\int_{\tau_0}^{\tau_1}\bigg[-\delta x^\mu g_{\mu\nu}\frac{D
\dot{x}^\nu}{D\tau}+\frac{d}{d\tau}\bigg(\delta x^\mu p_\mu \bigg)\bigg]d\tau\:, $$ with \be\lb{momo}
p_\mu=\frac{\delta \textit{L}}{\delta \dot{x}^\mu}=g_{\mu\nu}\dot{x}^\nu, \ee the conjugated moment of the
particle. For variations with fixed endpoints the total derivative in (\ref{lagro2}) can be discarded. The
variation will then vanish if and only if the equations of motion \be\lb{ekmok} \frac{D
\dot{x}^\mu}{D\tau}=\ddot{x}^\mu+\Gamma^\mu_{\nu\alpha}\dot{x}^\nu \dot{x}^\alpha=0 \ee are satisfied. Here
$\Gamma^\mu_{\nu\alpha}$ denotes the usual Christoffel symbols constructed in terms of the metric $g_{\mu\nu}$
\be\lb{crist} \Gamma_{ij}^k=\frac{g^{kl}}{2}(g_{il,j}+g_{jl,i}-g_{ij,l}), \ee and the first two terms of
(\ref{ekmok}) are simply the definition of the derivative $\frac{D \dot{x}^\nu}{D\tau}$. The system of equations
(\ref{ekmok}) states that the free particle in the geometry moves along a geodesic.

Consider now variations $\delta x^\mu=K^\mu$ without fixed endpoints. In this situation the time derivative can
not be ignored. By taking into account the equations of motion (\ref{ekmok}) it follows that \be\lb{sum} \delta
S=\int_{\tau_0}^{\tau_1}\delta \textit{L}\,\,d\tau=\int_{\tau_0}^{\tau_1}\frac{d}{d\tau}\bigg(K^\mu p_\mu
\bigg)d\tau. \ee If in addition $\delta x^\mu=K^\mu$ is such that $\delta S$ is zero, then this transformation
is a local symmetry of $\textit{L}$. From (\ref{sum}) it follows that the quantity \be\lb{shro} E_K= K_\mu
\dot{x}^\mu, \ee is a constant of motion associated to the symmetry. Thus, there exist a constant of motion for
every symmetry the lagrangian (\ref{lagro}) admits. A well known example of symmetries are the usual isometries,
which corresponds to local variations of the form $\delta x^\mu=K^\mu(x)$ which leave the action invariant. For
them, the vanishing of (\ref{sum}) gives that \be \frac{d }{d\tau}\bigg(K_\mu \dot{x}^\mu \bigg)
=\dot{x}^\nu\nabla_\nu K_\mu \dot{x}^\mu+K_\mu \frac{D \dot{x}^\mu}{D\tau}=0. \ee But since the first term is
zero due to (\ref{ekmok}) it follows that \be\lb{killing} \nabla_{(\nu} K_{\mu)}=0, \ee is satisfied for the
generators of the isometry. Here the parenthesis denote the symmetrization operation. The vectors satisfying
(\ref{killing}) are known as Killing vectors and are by definition the generators of the isometries.
Nevertheless, the isometries are not the most general symmetries. One may consider for instance transformations
of the form $\delta x^\mu=K(x,\dot{x})$, which are local in the phase space ($x^\mu$, $\dot{x}^\mu$). This form
is quite general, since any dependence in higher order time derivatives such that $\ddot{x}$ will be reduced to
a combination of ($x$, $\dot{x}$) by the equations of motion (\ref{ekmok}). Given such a symmetry one may impose
a Taylor like expansion of the form \be\lb{expans} \delta x^\mu=K^\mu+K^\mu_\alpha
\dot{x}^\alpha+K^\mu_{\alpha\beta} \dot{x}^\alpha \dot{x}^\beta+..., \ee with tensors $K^\mu_{\mu_1..\mu_n}(x)$
independent of the velocities $\dot{x}_i$. In these terms, it may be shown that (\ref{expans}) is a symmetry of
(\ref{lagro}) when \be\lb{killingt} \nabla_{(\mu} K_{\mu_1..\mu_n)}=0, \ee is satisfied. The reasoning for
reaching this conclusion is completely analogous to the one giving (\ref{killing}) and in fact (\ref{killingt})
are a generalization of the Killing condition for symmetric tensors of higher order. The tensors satisfying that
condition are known as \textit{Killing tensors} and the quantities \be\lb{canti}
C_n=K_{\mu_1..\mu_n}\dot{x}^{\mu_1}..\dot{x}^{\mu_n}, \ee are all constants of motion.

The constants (\ref{canti}) are a homogeneous polynomial expressions in the velocities whose degree is equal to
the order of the associated Killing tensor. For tensors of order larger than one, the corresponding symmetries
are not as visual or intuitive as usual isometries. For this reason these are known as �hidden symmetries�.

There exist another important generalization of the Killing vectors, which are the Killing-Yano tensors
\cite{Yano}. Although several of their properties has been reviewed for instance in \cite{Van Holten lectures}
we briefly state their main properties. Killing vectors and Killing tensors generate scalar constants of motion
(\ref{canti}). This scalar quantity of course takes the same value along the geodesic where the particle moves.
If instead tensor "conserved" quantities are considered, one should compare its components in different points
on the manifolds. This requires parallel transport. The statement that a tensor quantity $C_{\mu_1...\mu_{n-1}}$
is conserved then means that it is parallel transported along the geodesic. This is satisfied when \be\lb{mv}
\dot{x}^{\alpha}\nabla_{\alpha}C_{\mu_1...\mu_{n-1}}=0. \ee A Killing-Yano tensor $f_{\mu_1...\mu_n}$ is an
\emph{antisymmetric} one and which generate tensor "constants" of motion linear in the velocities \be
C_{\mu_1...\mu_{n-1}}=f_{\mu_1...\mu_{n-1}\mu} \dot{x}^{\mu}. \ee The parallel transport condition (\ref{mv})
implies that \be\lb{kiya} f_{\mu_1...\mu_{n-1}(\alpha;\beta)}=0. \ee This is a generalization of the Killing
vector equation for antisymmetric tensors of higher order. The square \be
C^2=C_{\mu_1...\mu_{n-1}}C^{\mu_1...\mu_{n-1}}=f_{\mu_1...\mu_{n-1}\mu} f^{\mu_1...\mu_{n-1}}_{\nu}
\dot{x}^{\mu}\dot{x}^{\nu} \ee is obviously a constant of motion quadratic in the velocities. This means that
the "square" \be K_{\mu\nu}=f_{\mu_1...\mu_{n-1}\mu} f^{\mu_1...\mu_{n-1}}_{\nu}, \ee is a Killing tensor of
order two. This result is usually paraphrased by saying that the square of a Killing-Yano tensor
$f_{\mu_1...\mu_{n-1}\mu}$ of arbitrary order is a Killing tensor $K_{\mu\nu}$ of order two. This  statement is
also true when the Killing tensor is constructed out of two different Killing-Yano tensors as follows $$
K_{\mu\nu}=f^{(1)}_{\mu_1...\mu_{n-1}(\mu} f^{(2)\mu_1...\mu_{n-1}}_{\nu)}, $$ but this will not be relevant in
the following analysis.

\subsection{A new cuadratic constant for the $Y(p,q)$ geometries}

In the present section a further quadratic constant of motion for the $Y(p,q)$ manifolds will be introduced. The
technical details of the construction are described in the Appendix but their validity will be shown below.

Our main statement is that the following tensor \be\lb{Killing} K_{\mu\nu}= \left( \begin{array}{ccccc}
K_{\theta\theta} & 0 & 0 & 0 &0\\ 0 & K_{\phi\phi} & 0 & K_{\phi\beta} & K_{\phi \psi'}\\ 0 & 0 & K_{y y} & 0 &
0\\ 0 & K_{\beta\phi} & 0 & K_{\beta \beta} & K_{\beta \psi'}\\ 0 & K_{\psi' \phi} & 0 & K_{\psi' \beta} &
K_{\psi' \psi'} \end{array} \right) \ee whose components are explicitly \be \begin{array}{l}
K_{\theta\theta}=\frac{4}{3}(1-y)\\\\
K_{\phi\phi}=\frac{1}{9}\left\{[1+\cos(2\theta)][q(y)w(y)+8y^2]+\cos(2\theta)[2-10y]+14-22y\right\}\\\\ K_{y
y}=\frac{8}{q(y)w(y)} \\\\ K_{\beta \beta}=\frac{2}{9}[q(y)w(y)+8y^2]\\\\ K_{\psi' \psi'}=\frac{16}{9}\\\\
K_{\phi\beta}=K_{\beta\phi}=\frac{2}{9}[q(y)w(y)+8y^2-8y]\cos(\theta)\\\\ K_{\phi \psi'}=K_{\psi
\phi}=\frac{16}{9}(y-1)\cos(\theta)\\\\ K_{\beta \psi'}=K_{\psi' \beta}=\frac{16}{9}y, \end{array} \ee is a
Killing one. These assertion was checked with the help of the Ricci package of the Wolfram Mathematica program,
which gives as a result that $K_{(\mu\nu;\lambda)}=0$ for every choice of indices. This means that the
infinitesimal transformation $\delta\dot{x}^{\mu}=K^{\mu}_{\alpha}\dot{x}^{\alpha}$ represents a hidden symmetry
for the geodesic motion in the $Y(p,q)$ Einstein-Sasaki metric (\ref{ypq}). The associated constant of motion is
\be\lb{constante'}
C=K_{\mu\nu}\dot{x}^{\nu}\dot{x}^{\mu}=K_{\theta\theta}\dot{\theta}^2+K_{\phi\phi}\dot{\phi}^2+K_{y
y}\dot{y}^2+K_{\beta \beta}\dot{\beta}^2+K_{\psi' \psi'}\dot{\psi'}^2 \ee $$
+2K_{\phi\beta}\dot{\phi}\dot{\beta}+2K_{\phi \psi'}\dot{\phi}\dot{\psi'}+2K_{\beta
\psi'}\dot{\beta}\dot{\psi'}. $$ By going to the coordinates $(\theta,\phi,y,\alpha,\psi)$ which takes the
metric to the form (\ref{metric2}) the constant may be expressed as follows $$
C=\frac{4}{3}(1-y)\dot{\theta}^2+\frac{8}{q(y)w(y)}\dot{y}^2+8\left[q(y)w(y)+8y^2\right]\dot{\alpha}^2+\left\{16\left[q(y)w(y)+8y^2\right]-\frac{16}{3}\right\}\dot{\alpha}\dot{\psi}
$$ $$
+\frac{1}{9}\left\{\left[1+\cos\left(2\theta\right)\right]\left[q(y)w(y)+8y^2\right]+\cos(2\theta)\left[2-10y\right]+14-22y\right\}\dot{\phi}^2-\frac{24}{9}\left[q(y)w(y)\right.
$$ $$
\left.+8y^2-8y\right]\cos(\theta)\dot{\phi}\dot{\alpha}+\left\{\frac{32}{9}(y-1)\cos\theta-\frac{24}{9}\left[q(y)w(y)+8y^2-8y\right]\cos(\theta)\right\}\dot{\phi}\dot{\psi}
$$ \be\lb{constanteexpl} +\left\{\frac{16}{9}-\frac{64}{3}y+8\left[q(y)w(y)+8y^2\right]\right\}\dot{\psi}^2. \ee
The results presented above are not enough to state that the geodesic equations on $Y(p,q)$ are superintegrable.
This will be the case if if the set $\left\{P_\phi, P_\psi, P_\alpha, J^2, H, C\right\}$ constitute a
functionally independent set of constants of motion for the massless geodesics on $AdS_5\times Y(p,q)$ geometry
considered in previous sections.  To prove the functional independence one should construct the $(d+1) \times
2d$ Jacobian \be\lb{jacob}
J=\frac{\partial(P_\phi,P_\psi,P_\alpha,J^2,H,C)}{\partial(\theta,\phi,y,\alpha,\psi,\dot{\theta},\dot{\phi},\dot{y},
\dot{\alpha},\dot{\psi})} \ee with $d=5$ and to calculate its rank. The result is \be\lb{rango} Rank(J)=6, \ee
and was also obtained by use of the Wolfram Mathematica program. Therefore it is safe to say that the
configuration of massless geodesics on $AdS_5\times Y(p,q)$ defined in previous sections are superintegrable,
since the number of degrees of freedom is five and the number of functionally independent constants of motion is
at least six.

\section{Comparison with the literature} The purpose of this section is to derive some features related to the
presence of the quadratic constant (\ref{constanteexpl}) for the Einstein-Sasaki $Y(p,q)$ geometries. Our
analysis will rely in some standard results in the literature, which we will cite explicitly below.

\subsection{Separability of the Laplace and the Dirac operators} The Killing tensor (\ref{constanteexpl}) is the
square of a Killing-Yano 3-form $K_{\mu\nu}=f_{\mu\alpha\beta}f_{\nu}^{\alpha\beta}$ (see Appendix). The
explicit expression of this 3-form is
\be\lb{ESKYexp} f=\frac{1}{9}\;[(1-y)\sin\theta \;d\theta\wedge
d\phi\wedge d\psi' - \cos\theta \;d\phi\wedge dy\wedge d\psi'+ dy \wedge d\beta\wedge d\psi'
\ee
$$ + \, \sin\theta \;y(1-y)\; d\theta\wedge d\phi\wedge d\beta - \cos\theta \, d\phi\wedge dy\wedge d\beta], $$

The first implicance
is that the hidden symmetry that the Killing tensor (\ref{constanteexpl}) generates is not anomalous. This
statement may be explained as follows. The quantum mechanical analog of the hamiltonian for the free particle in
a curved geometry is the laplacian $$ \Delta=\nabla_{\mu}(g^{\mu\nu}\nabla_{\nu}). $$ Any Killing vector
$V=V^{\mu}\partial_{\mu}$ for $g_{\mu\nu}$ is in correspondence with a quantum mechanical operator
$\hat{V}=V^{\mu}\nabla_{\mu}$ which commutes with the laplacian $\Delta$ defined above. But this is not the case
for Killing tensors \cite{Carter}, unless some extra conditions are satisfied. In fact, the commutator of the
operator $\hat{K}=\nabla_{\mu}(K^{\mu\nu}\nabla_{\nu})$ associated to the Killing tensor $K_{\mu\nu}$ with the
laplacian $\Delta$ is given by \cite{Carter} $$ [\widehat{K}, \widehat{H}]=
-\frac{4}{3}\nabla_{\nu}(R_{\mu}^{[\nu} K^{\sigma]\mu})\nabla_{\sigma}, $$ which is not zero in general. This
means that the symmetry a Killing tensor generates is anomalous unless the integrability condition
\be\lb{piaza}R_i^{[j} K^{k]i}=0,\ee is satisfied. This condition holds when the space is Einstein
$R_{ij}=\Lambda g_{ij}$, or when the Killing tensor is the square of a Killing-Yano tensor
\cite{Yano}-\cite{WalkerPenrose}. Both conditions are satisfied for the $Y(p,q)$ geometries, therefore the
hidden symmetry that (\ref{killing}) generates is \emph{not} anomalous.

The presence of a Killing-Yano tensor like (\ref{ESKYexp}) is also relevant for studying the separability of the
Dirac operator in the geometry \cite{Cariglia}-\cite{Cariglia2}. For geometries admiting spinors one can
consider an irreducible representation of the Clifford algebra, which is composed by elements $e^a$ for which
\be e^a e^b+e^b e^a=g^{ab}. \ee The Dirac operator in the geometry is \be\lb{dirc} D=e^a \nabla_{X_a}. \ee Given
an arbitrary $p$-form $f_{\mu_1...\mu_p}$ one may construct an operator $D_f$ with special properties. It is
given explicitly as \be\lb{volv} D_f=L_f-(-1)^p f D, \ee with \be
L_f=e^af\nabla_{X^a}+\frac{p}{p+1}df+\frac{n-p}{n-p+1}d^\ast f. \ee By defining the graded commutator \be \{D,
D_f\}=D D_f+(-1)^p D_f D, \ee it follows from (\ref{volv}) and (\ref{dirc}) that \be\lb{sevvv} \{D, D_f\}=R D,
\qquad R=\frac{2(-1)^p}{n-p+1}d^{\ast} f D. \ee For a Killing-Yano $p$-form $f_{\mu_1...\mu_p}$ it may be shown
from the definition (\ref{kiya}) that $d^{\ast}f=0$ and therefore $R=0$. By comparing this with (\ref{sevvv}) it
follows immediately that for any Killing-Yano tensor $f_{\mu_1...\mu_p}$ of arbitrary order there exist an
operator $D_f$ acting on spinors and whose graded commutator with the Dirac operator $D$ is zero
\cite{Cariglia}. We remark that for $p=3$, as in (\ref{ESKYexp}), the graded commutator becomes the usual
commutator.

\subsection{Exotic supersymmetries}

In addition to the applications described in the previous subsection, the presence of a Killing-Yano tensors in
a manifold $M$ with metric $g_{\mu\nu}$ has applications related to the worldline action for the superparticle
described in \cite{Superparticula1}-\cite{Superparticula4}. This action can be writen in terms of a superfield
$X$ which maps a supermanifold parameterized by two coordinates $(t,\theta)$ into $M$. The variable $\theta$ is
a Grassman variable, which means that $\theta^2=0$. The worldline action is $$ I=\frac{i}{2}\int dt d\theta
g_{\mu\nu}D^{\mu}\partial_t X^{\nu}, $$ and it is supersymmetric by construction. Here $D$ is the worldline
superspace derivate, for which $D^2=i\partial_t$. In addition to the usual supersymmetries, if the space-time
metric $g_{\mu\nu}$ admits a set Killing-Yano tensors $f^i_{\mu_1...\mu_p}$ there appear new symmetries for the
action of the form \cite{Papadopoulos1} $$ \delta X^{\mu}=\epsilon_{i}f^i_{\mu_1...\mu_p}DX^{\mu_1}..DX^{\mu_p},
$$ with $\epsilon_i$ infinitesimal parameters. These symmetries imitate the supersymmetry property of mixing the
bosonic and fermionic components of the superfield $X$, but their algebra is not the supersymmetry algebra.
Additionally, they are generated by space-time forms. For this reason they are sometimes known as "exotic
supersymmetries" \cite{susyinthesky}-\cite{susyinthesky2}. Thus we conclude from (\ref{ESKYexp}) that the
superparticle worldline action defined on the $Y(p,q)$ geometries admits at least one exotic supersymmetry.

\section{Discussion and open perspectives}

In the present work a Carter like constant for the $Y(p,q)$ Einstein-Sasaki metrics was constructed. The
constant we found is functionally independent with respect to the five known constant of motion for the
geometry. The complete set of functionally independent constants for the problem is at least six and since there
are five degrees of freedom, the geodesic equations turns out to be superintegrable.

It should be emphasized that our constant of motion is constructed in terms of the square of a Killing-Yano
tensor. The standard results found by Carter and studied deeply in \cite{Yano}-\cite{WalkerPenrose} insures that
this constant of motion is in correspondence with an operator that commutes with the laplacian. In other words,
the hidden symmetry we have constructed is not anomalous.

We also pointed out that the worldline superparticle action \cite{Superparticula1}-\cite{Superparticula4} over
the $Y(p,q)$ geometries admits an exotic supersymmetry. Aside from that, it may be an interesting general task
to realize whether or not the fake supersymmetries found in \cite{van riet} may be interpreted as exotic
supersymmetries.

We ignore if the Killing-Yano tensor (\ref{ESKYexp}) and the Killing tensor (\ref{Killing}) we constructed were
presented before. By looking the literature we have in hand we suggest that the present work may overlap with
\cite{Yasui}. In that reference the spectrum of certain BPS operators for a gauge theory dual to IIB
superstrings in a geometry of the form $AdS_5$ x \textit{(Einstein-Sasaki)} was studied. This spectrum can be
analized in terms of the eigenfunctions of the Laplacian defined over the Einstein-Sasaki internal space. The
authors of that reference are able to separate the eigenvalue equations completely and they found an accidental
quadratic constant of motion \textit{C} during the process. There are four constants of motion for these
geometries, which are related to usual isometries and the energy of the configuration. The constant \textit{C}
instead is not related to an usual isometry. But since these geometries are obtained by reduction of certain
black hole geometries which admit Killing tensors, these authors suggest that \textit{ C} may be related to the
hidden symmetries of the higher dimensional geometry.

The reference described above is related to a family of Einstein-Sasaki geometries which reduce to the $Y(p,q)$
ones in certain limit of their moduli. We are not completely sure at the moment if the constant \textit{C} found
in that reference will reduce to ours after taking this limit. If this is not the case, then the $Y(p,q)$
geometries admit two hidden symmetries instead of one. This is a possibility that deserves further attention.
But even if they coincide, the construction of the tensors (\ref{ESKYexp}) and (\ref{constanteexpl}) we are
doing is intrinsic, without taking into account reductions comming from higher dimensional black holes.
Additionally, we are pointing out that any Einstein-Sasaki geometry admits hidden symmetries (see appendix). For
this reason, even if we would rediscovered some results of that reference, we are using different arguments
which can be applied to more general situations \cite{giataganas}. \\

{\bf Acknowledgements:}  Discussions with I. Dotti, M. L. Barberis, G. Giribet, E. Eiroa, L. Chimento, J. Russo
and H. Vucetich are warmly acknowledged. O. S was benefited with discussions with F. Cuckierman about contact
structures. Both authors are supported CONICET (Argentina) and O. S is also supported by the ANPCyT grant
PICT-2007-00849. \\

\appendix \section{Sasaki and Einstein-Sasaki metrics} \subsection{Defining equations for Sasaki estructures}
Let us consider the following conical metric in six dimensions \be\lb{conus} g_{6}=dr^2+r^2 g_{5}. \ee Here the
metric $g_{5}$ does not depend on the radial coordinate $r$ and is defined over a variety which we denote as
$M_5$. The metric $g_6$ is defined over $R_{>0}\times M_5$ and is singular at the tip of the cone $r=0$. The
metric $g_{5}$ is known as Sasaki if the cone $g_{6}$ is Kahler. The converse also holds. Let us recall that a
metric $g_6$ is Kahler if there exist a basis $e^a$ such that $$ g_6=\delta_{ab}e^a\otimes e^b, $$ and  such
that the almost complex structure \be\lb{cc} J=e_1\otimes e^2-e_2\otimes e^1+e_{3}\otimes e^{4}-  e_{4}\otimes
e^{3}+e_{5}\otimes e^{6}-e_{6}\otimes e^{5}, \ee is covariantly constant, i.e, $\nabla_X J=0$ for any vector
field $X$ in $TM_6$. This condition implies that the manifold is complex and sympletic with respect to the two
form $\omega=g(\cdot, J\cdot)$. The sympletic form is explicitly $$ \omega=e^1\wedge e^2+e^3\wedge e^4
+e^5\wedge e^6. $$ If furthermore $g_6$ is Ricci flat, then $g_6$ is Einstein $g_5$ and is usually denominated
as Einstein-Sasaki. A Ricci flat Kahler metric has an holonomy group included in $SU(3)$ and is called
Calabi-Yau. Thus there is a one to one correspondence between the non compact Calabi-Yau cones (\ref{conus}) and
Einstein-Sasaki manifolds.

It is convenient for the following discussion to select a basis for $g_6$ of the form \be\lb{bas} e_i=r
\widetilde{e}_i,\qquad e_6=dr, \ee with $i=1,..,5$ y $\widetilde{e}_i$ a basis for $g_{5}$. In this basis
(\ref{bas}) the almost complex structure (\ref{cc}) takes the form \be J=\widetilde{e}_1\otimes
\widetilde{e}^2-\widetilde{e}_2\otimes \widetilde{e}^1+\widetilde{e}_{3}\otimes
\widetilde{e}^{4}-\widetilde{e}_{4}\otimes \widetilde{e}^{3}+r \widetilde{e}_5\otimes
\partial_r-\frac{dr}{r}\otimes \widetilde{e}^5. \ee Alternatively, it may be expressed as \be\lb{almcom}
J=\phi+r \eta\otimes \partial_r-\frac{dr}{r}\otimes \xi \ee with $\eta=e_5$ and $\xi=e^5$, which means that
$\eta(\xi)=1$. The quantity $\phi$ is defined as \be\lb{fifi} \phi=\widetilde{e}_1\otimes
\widetilde{e}^2-\widetilde{e}_2\otimes \widetilde{e}^1+\widetilde{e}_{3}\otimes
\widetilde{e}^{4}-\widetilde{e}_{4}\otimes \widetilde{e}^{3}. \ee The expression (\ref{fifi}) for $\phi$
involves four elements of the basis, but this does not mean that $\phi$ is a quantity defined in a subvariety
four dimensional $M_4$ of $M_5$. In fact the elements $\widetilde{e}^a$ with $a=1,2,3,4$ are 1-forms defined
over $M_5$. Nevertheless, as we will discuss now, if the metric is Kahler then these elements are defined on a
subvariety $M_4$, and $\phi$ becomes an almost complex structure for $M_4$. This can be checked as follows. A
vector field $\widetilde{X}\in R_{>0}\times M_{5}$ may be decomposed in a radial part $a$ and an angular part
$X$ such that $\widetilde{X}=(a, X)$. Starting from (\ref{almcom}) it may be deduced that the action of the
almost complex structure over $\widetilde{X}$ is given as \be\lb{axion} J(a, X)=(r\eta(X), \phi
X-\frac{a}{r}\xi). \ee Furthermore, the Levi-Civita connection $\widetilde{\nabla}$ for the cone may be
decomposed in the following way $$ \widetilde{\nabla}_{\partial_r}\partial_r=0,\qquad
\widetilde{\nabla}_{X}\partial_r=\widetilde{\nabla}_{\partial_r}X=\frac{X}{r} $$ \be\lb{impous}
\widetilde{\nabla}_X Y=\nabla_X Y-r g(X,Y)\partial_r. \ee Here $\nabla$ is the conection for $g_{5}$. Formula
(\ref{impous}) is completely elementary and arise directly by comparing the Christofell symbols of $g_5$ with
the ones for $g_6$. From (\ref{impous}) and (\ref{almcom}) it is obtained the following action in this basis $$
(\widetilde{\nabla}_{\partial_r} J)\partial_r=(0,0),\qquad (\widetilde{\nabla}_{\partial_r}) X=(0,0), $$
\be\lb{simama} (\nabla_X J)\partial_r=(0,\frac{1}{r}(-\nabla_X \xi+\phi X)) \ee $$ (\widetilde{\nabla}_{X}
J)Y=(r \nabla_X \eta(Y)-r g_{5}(X,\phi Y), (\nabla_X\phi)Y-g_{5}(X,Y)\xi+\eta(Y)X). $$ As we remarked above the
cone $g_6$ will be Kahler if $\nabla_X J=0$, which means that all the covariant derivatives (\ref{simama})
should be zero. This happens if and only if \be\lb{ES1} \nabla_X \xi=\phi X, \ee \be\lb{ES2} \nabla_X
\eta(Y)=g_{5}(X,\phi Y), \ee \be\lb{ES3} (\nabla_X\phi)Y=g_{5}(X,Y)\xi-\eta(Y)X. \ee The radial coordinate $r$
does not appear in these expressions and thus these are constraints on $g_5$. The metrics $g_5$ which satisfy
(\ref{ES1}), (\ref{ES3}) and (\ref{ES3}) are Sasakian by definition.

\subsection{Derivation of the main formulas (\ref{Killing}) and (\ref{ESKYexp})  of the text}

The relations (\ref{ES1})-(\ref{ES3}) as written above may not be very illuminating, but they may be clarified
by examining their consequences. The relation (\ref{ES2}) implies that \be
\nabla_X\eta(Y)+\nabla_Y\eta(X)=g_{5}(X,\phi Y)+g_{5}(Y,\phi X), \ee but it may be directly deduced from the
definition (\ref{fifi}) of $\phi$ that $g_{5}(X,\phi Y)=-g_{5}(Y,\phi X)$. Thus our last equation is
\be\lb{cul}
 \eta_{(i;j)}=0,
\ee which implies that $\xi=\eta^{\ast}$ is a Killing vector and we have the local decomposition
$M_5=U_{\xi}(1)\times M_4$ as anticipated. The metric takes the form $$ g_5=\eta^2+g_4, $$ with $$
g_4=\widetilde{e}^1\otimes\widetilde{e}^1+ \widetilde{e}^2\otimes\widetilde{e}^2+
\widetilde{e}^3\otimes\widetilde{e}^3 +\widetilde{e}^4\otimes\widetilde{e}^4. $$ This is the local form
(\ref{loco}) presented in the text. The vector field $\xi=\eta^{\ast}$ is the Reeb vector. Additionally if we
define $I$ the restriction of $\phi$ to $M_4$ then it follows from (\ref{ES2}) and (\ref{cul}) that $$ g_4(u, I
v)=d\eta(u,v), $$ for arbitrary vectors $u$ and $v$ in $TM_4$. Denoting \be\lb{uta} f(u,v)=d\eta(u,v) \ee it
follows that $d_4f=0$, thus $M_4$ is sympletic. In addition the antisymmetric part of (\ref{ES3}) can be written
by taking into account (\ref{ES1}) and uppering indices with the metric $g_5$ as follows \cite{Sewemann2}
\be\lb{saskon} \nabla_X(d\eta)=-2 X^{\ast}\wedge \eta. \ee Note that for vectors $u$ in $TM_4$ the right hand
side of (\ref{saskon}) is zero, since vector fields in $TM_4$ are orthogonal to $\eta$. This means that $$
\nabla_4 d\eta=\nabla_4 f=0, $$ where we took into account the definition (\ref{uta}). The last condition says
that the metric $g_4$ defined on $M_4$ is Kahler. This one of the features discussed below formula (\ref{ke}) in
the text.

Finally, let us explain how the Killing-Yano tensor (\ref{ESKYexp}) and the Killing tensor (\ref{Killing}) were
obtained. For this purpose, consider the definition of Killing-Yano tensors (\ref{kiya}). This definition is
completely equivalent to the following \be\lb{cherri} \nabla_X f=\frac{1}{p+1}i_X df, \ee with $X$ an arbitrary
vector field and $i_X$ the usual contraction operation. The equivalence can  be checked directly by writting
(\ref{cherri}) in components. The conformal generalization of (\ref{cherri}) was found in \cite{Conformal
KY}-\cite{Curto}, and is given by \be\lb{cucho} \nabla_X f=\frac{1}{p+1}i_X df-\frac{1}{n-p+1}X^{\ast}\wedge
d^{\ast}f. \ee Here $X^{\ast}$ is the dual 1-form to the vector field $X$ and the operation $d^{\ast}
f=(-1)^p\ast^{-1} d \ast f$ has been introduced, in which $$ \ast^{-1}=\epsilon_p \ast,\qquad
\epsilon_p=(-1)^p\frac{\det g}{|\det g|}. $$ The p-forms satisfying the condition (\ref{cucho}) are known as
conformal Killing forms. When $d\ast f=0$, (\ref{cucho}) reduces to the usual definition of a Killing tensor
(\ref{cherri}).

Consider again (\ref{saskon}). A direct consequence of this condition is $d^{\ast}d\eta=2(n-1)\eta$, with $n=5$.
This means that (\ref{saskon}) can be expressed alternatively as \be\lb{saskon2} \nabla_X(d\eta)=-\frac{1}{n-1}
X^{\ast}\wedge d^{\ast}d\eta. \ee By taking into account that $d\eta$  is closed and denoting $f=d\eta$, then
comparison of (\ref{saskon2}) with (\ref{cucho}) shows that $d\eta$ is a conformal Killing tensor. This,
together with the fact that $\eta$ is a Killing 1-form implies that the combinations \be\lb{cob}
\omega_k=\eta\wedge (d\eta)^k, \ee are all Killing tensors of order $2k+1$. This calculation is straightforward
and we learned it from proposition 3.4 of \cite{Sewemann2}. Note that (\ref{cob}) is a \emph{generic} statement
for all the Sasakian structures. For the $Y(p,q)$ geometries, we have from (\ref{loco}) and (\ref{1f}) that
\be\lb{locol} \eta=d\psi^{'}+A=d\psi^{'}-\cos\theta d\phi+y(d\beta+\cos\theta d\phi), \ee and the tensor
(\ref{cob}) constructed with this form with $k=1$ results
\be\lb{ESKYexp2}
\omega_3=\frac{1}{9}\;[(1-y)\sin\theta \;d\theta\wedge d\phi\wedge d\psi' - \cos\theta \;d\phi\wedge dy\wedge d\psi'+ dy \wedge d\beta\wedge d\psi'
\ee
$$
+ \, \sin\theta \;y(1-y)\; d\theta\wedge d\phi\wedge d\beta - \cos\theta \, d\phi\wedge dy\wedge d\beta].
$$
which is the Killing-Yano tensor presented in formula (\ref{ESKYexp}) of the
text, up to a notational identification $\omega_3=f$. The square
$K_{\mu\nu}=(\omega_3)_{\mu\alpha\beta}(\omega_3)_{\nu}^{\alpha\beta}$ is the Killing tensor (\ref{Killing}) of
section 3. These are the two fundamental formulas utilized along the text. Note that the Einstein condition do
not play any role in obtaining these results, and can be generalized to other Einstein-Sasaki configurations
such as the ones studied in \cite{giataganas}.

 \end{document}